\documentclass[12pt]{article}
\usepackage{amsmath}
\usepackage{amsfonts}

\newcommand{\bean}{\begin{eqnarray*}}
\newcommand{\eean}{\end{eqnarray*}}
\newcommand{\ed}{\end{document}}

\newcommand{\be}{\begin{equation}}
\newcommand{\ee}{\end{equation}}
\newcommand{\barr}{\begin{array}}
\newcommand{\earr}{\end{array}}
\newcommand{\bea}{\begin{eqnarray}}
\newcommand{\eea}{\end{eqnarray}}

\begin{document}
\title{
Generalized inversion of 
the Hochschild  coboundary operator 
and deformation quantization.}
\author{A.V.Bratchikov
 \\ Kuban
State Technological University,\\ 2 Moskovskaya Street, Krasnodar,
  350072, Russia
\\
}
 \date{} \maketitle
\begin{abstract}
Using a derivative decomposition of the Hochschild differential complex we define a generalized inverse of the  Hochschild coboundary operator. It can be applied for systematic computations of star products on Poisson manifolds.
\end{abstract}



\section  {Introduction}
Let $M$ be a phase space with the phase variables $x_i,\,i=1,\ldots,n,$
and the Poisson bracket
\bean \label{U}
\{x_i,x_j
\}=
\omega_{ij}(x).
\eean
Let $L$ be a set of the multi-indeces
$a = (a^1, a^2,\ldots, a^n ),$
$|a |=a^1 + a^2+ \ldots +~a^n 
.$  
For $f\in A=C^\infty(X)$ we shall denote
\bean
X^{a}f(x)=\frac 1 {a^1!a^2!\ldots a^n!}\frac {\partial ^{|a|} f(x_1,x_2,\ldots, x_n)}
{\partial {x_1^{a^1}} \partial x_2^{a^2}\ldots \partial x_n^{a^n}}.
\eean

Let $\tilde A=A[[t]]$
be the space of formal power series in
a variable $t$
with coefficients in $A.$
A star product
is a $R[[t]]$ linear associative product on
$\tilde A,$  defined for $f,g\in A$ by
\bea \label {ore}
f\ast g=fg
+\sum_{k=1}^\infty t^k
\Pi^{k}
(f,
g)
,
\eea 
where $\Pi^k(f,g)$ are bidifferential operators: 
\bean \Pi^k(f,g)= X^{a}f \Pi^k_{a b}X^{b}g ,\quad \Pi^k_{a b}(x)\in A.
\eean
Functions $\Pi^k_{a b}$ are defined by the 
associativity equation  
\bea
\label {asse}
(f\ast g)\ast h=
f\ast (g\ast h)
\eea
for all
$f,g,h\in \tilde A.$

One can show that
\bea \label{one} \Pi^1(f,g)= 
\frac 1 2 X^{a}f \omega_{a b}X^{b}g
,
\quad a, b \in L_0,
\eea
where  $L_0$ is the set of unit vectors $\left \{e_1,e_2,\ldots, e_n \right\},\,e_i^j= \delta_i^j,$
and $ \omega_{e_i e_j}= \omega_{ij}.$

An explicit formula for
the star product  on an
arbitrary Poisson manifold was found by Kontsevich \cite {K}. It is given by a series of diagrams. However there is no systematic way to compute weights of the corresponding graphs.  
The cochains of the star product which define a third order deformation in variable $t$ were found in \cite {PV}. A second order deformation
in powers of derivatives of  Poisson structures is defined by the  Baker-Campbell-Hausdorff  formula  
for Poisson  bracket algebras \cite {B}. 
For symplectic Poisson manifolds one can use Fedosov's
construction of deformation quantization \cite {F}.

In this paper we  present a new systematic method of computation of star products on arbitrary Poisson manifolds. It is based on a derivative decomposition of the Hochschild differential complex and  generalized inversion of the  Hochschild coboundary operator.

\section {A derivative decomposition of the Hochschild differential complex}
Let $C^p(A), p\ge 0, $ be the space of differential $p~-$cochains.
One may view $C^p(A)$ as a vector bundle $$\pi: C^p(A)\to M.$$ 
For every $x\in M$   
$$ X^{a_1}\otimes X^{a_2}\otimes \ldots \otimes X^{a_p},\quad  a_1,\ldots ,a_n\in L,$$ is a basis of the fiber  $\pi^{-1}(x).$ A general element $\Phi(x)\in \pi^{-1}(x)$ reads   
\bean
\Phi(x) = \Phi_{a_1a_2\ldots a_p}(x)
X^{a_1}\otimes X^{a_2}\otimes \ldots \otimes X^{a_p}
\eean
where $\Phi_{a_1a_2\ldots a_p}\in A.$ 
For $f_1,f_2,\ldots, f_p \in A$
\bean
\Phi(f_1,f_2,\ldots, f_{p}) = \Phi_{a_1a_2\ldots a_p}X^{a_1}f_1 X^{a_2}f_2 \ldots  X^{a_p}f_p.
\eean

Let ${\bf \delta}$ be the Hochschild coboundary operator 
\bean
\delta: C^p(A)\to C^{p+1}(A)
\eean
which is defined by
\begin {multline} 
\label {nutu} 
\delta \Phi (f_1,f_2,\ldots, f_{p+1})= f_1\Phi (f_2,\ldots, f_{p+1})+ \\
+\sum_{k=1}^p(-1)^k
\Phi (f_1,\ldots, f_{k-1},f_k f_{k+1},f_{k+2},\ldots, f_{p+1})+(-1)^{p+1}\Phi (f_1,\ldots, f_p)f_{p+1}.
\end {multline}
Definitions of $\pi$ and $\delta$ show that $\delta \pi = \pi \delta.$

Equation (\ref {nutu}) can be written in the form  
\begin {multline*} \delta \Phi (f_1,f_2,\ldots, f_{p+1})
= 
\sum_{k=1}^p(-1)^{(k+1)}
\bigl( f_k \Phi (f_1,\ldots, f_{k-1},f_{k+1},\ldots, f_{p+1})+\\
+f_{k+1}\Phi (f_1,\ldots,f_k,f_{k+2},\ldots, f_{p+1})-\Phi (f_1,\ldots, f_{k-1},f_k f_{k+1},f_{k+2},\ldots, f_{p+1})
\bigr)
.
\end
{multline*}
One finds 
\bean
\delta \Phi =  \Phi_{a_1\ldots a_p}\delta (X^{a_1}\otimes \ldots\otimes X^{a_p}),
\eean
\bean
\delta (X^{a_1}\otimes \ldots \otimes X^{a_ p})=\sum_{k=1}^p(-1)^{(k+1)}
X^{a_1}\otimes  \ldots \otimes\delta X^{a_k}\otimes\ldots \otimes X^{a_p}
.
\eean
Here
$$
\delta X^a = -\mathop{ {\sum}'}_{s=0}^a X^{a-s}\otimes X^s,$$ $$\mathop{ {\sum}'}_{s=0}^a Y^{s}= \sum_{s=0}^a Y^{s}- Y^{a} -Y^0,\qquad
\sum_{s=0}^a =\sum_{s_1=0}^{a_1}\ldots \sum_{s_p=0}^{a_p}.
$$

Let $A=(a_1,a_2,\ldots, a_p), B=(b_1,b_2,\ldots ,b_{p+1})$ be multi-indeces,$\Phi_A=\Phi_{a_1\ldots a_p}$ and $X^B=X^{b_1}\otimes \ldots \otimes X^{b_{p+1}}.$ Then 
\bean
\delta \Phi= \Phi_{A}\Delta^{A}_{B}X^{B},
\eean
where
\bea \label {matr}
\Delta^{A}_{B}=\sum_{k=1}^p (-1)^{k+1} \delta^{a_1}_{b_1}\ldots \delta^{a_{k-1}}_{b_{k-1}}Q^{a_k}_{b_k  b_{k+1}}\delta^{a_{k+1}}_{b_{k+2}}\ldots \delta^{a_{p}}_{b_{p+1}}
\eea
and 
\bean
Q^a_{b c}= 
-\delta^a_ {b +c}+\delta^0_ {b}\delta^a_ {c}+\delta^0_ {c}\delta^a_ {b},\qquad
\delta^b_a =  \delta^{b_1}_{a_1}\delta^{b_2}_{a_2}\ldots \delta^{b_n}_{a_n}.
\eean

Any space $C^p(A)$ can be uniquely decomposed as 
\bea \label {deca}
 C^p(A)= \bigoplus_{l\in L} C^{p}_l(A)
\eea
where $C^{p}_l(A) \cap \pi^{-1}(x
)$ is generated by $$X^{a_1}\otimes \ldots \otimes 
X^{a_p},\qquad
{a_1+\ldots+a_p=l}.
$$
The spaces $C^{p}_l(A)$ and $C^{p}_m(A), l\ne m,$ are orthogonal with respect to the fiberwise inner product
\bean \label {prod}
\left
<X^{a_1}\otimes\ldots \otimes X^{a_p},X^{b_1}\otimes\ldots \otimes X^{b_p}
\right
>= \delta_{a_1b_1} 
\ldots \delta_{a_p b_p}.
\eean

For $\Phi \in C^p_l(A)$ one gets
\bea \label {decoa}
{\delta}\Phi \in  C^{p+1}_l(A),
\eea
and therefore the Hochschild cochain complex
$$
C^\bullet(A):\quad\ldots \longrightarrow C^p(A)\longrightarrow C^{p+1}(A) \longrightarrow\ldots \phantom {\qquad l\in V.}
$$
splits into a direct sum of complexes 
$$
C^\bullet_l(A):\quad \ldots \longrightarrow C^p_l(A)\longrightarrow C^{p+1}_l(A) \longrightarrow\ldots \qquad l\in L.  
$$

The complex $C^\bullet_l(A)$ can be decomposed futher. 
Let  $\tilde C^p_l(A)$ be the subspace of $C^p_l(A)$ which is generated by 
\bean
X^{a_1}\otimes \ldots \otimes X^{a_p},\qquad a_i \ne 0,\quad i=1,\ldots, p            
\eean
and  $C^p_l(A)=\tilde C^p_l(A)\oplus \tilde C^{p \bot }_l(A).$
Then $\tilde C^{\bullet}_l(A)$ and $\tilde C^{\bullet \bot }_l(A)$ are the Hochschild complexes.

\section {Generalized inversion of $\delta$}
According to (\ref {deca}) and (\ref {decoa}) the  matrix  $\Delta= (\Delta^A_B)$ (\ref {matr})  can be decomposed as a direct sum of finite-dimensional
matrices
\bean \label {dect}
\Delta= \bigoplus_{l\in L} \Delta_l.
\eean

Let $\Delta^{{+}}_l$ be the Moore-Penrose generalized inverse of $\Delta_l.$ It is defined by the equations
\bean  \Delta^{\phantom{+}}_l\Delta^+_l \Delta^{\phantom {+}}_l=\Delta^{\phantom{+}}_l,\qquad
\Delta^{{+}}_l\Delta^{\phantom{+}}_l \Delta^{ {+}}_l=\Delta^{{+}}_l,
\eean
\bean\label {def}(\Delta
^{{+}}_l \Delta^{\phantom{+}}_l)^T= \Delta
^{{+}}_l \Delta^{\phantom{+}}_l,\qquad  (\Delta^{\phantom{+}}_l\Delta^{{+}}_l)^T=\Delta^{\phantom{+}}_l\Delta^{{+}}_l.
\eean
Here $u^T$ denotes the transpose of $u.$

Let $\delta^{{+}}:  C^{p+1}(A)\to C^{p}(A) $ be the operator which is represented by the matrix 
\bean \label {dec}
\Delta^{{+}}= \bigoplus_{l\in L} \Delta^{{+}}_l.
\eean
It is clear that 
\bean \label {ps} \delta\delta^+ \delta =\delta,\quad
\delta^{+}\delta \delta^{+}=\delta^{{+}}, 
\quad
(\delta
^{{+}} \delta)^T= \delta
^{{+}} \delta,\quad  (\delta\delta^{+})^T=\delta \delta^{+}.
\eean

An alternative  definition of $\delta^{{+}}$ is based on the Tihonov regularization 
\bea \label {tih}
\delta^+=
\lim_{\alpha \to 0}  \left(\alpha^2 I+ \delta^T \delta \right)^{-1}\delta^T
=\lim_{\alpha \to 0} \delta^T \left(\alpha^2 I+ \delta \delta^T \right)^{-1},
\eea
where $I$ is the identity operator. The operator $\delta^T: C^{p+1}(A)\to C^p(A)$ satisfies
\bean \label {trans}
\langle \delta^T \Psi,\Phi
\rangle
=
\langle \Psi,\delta \Phi
\rangle \qquad \mbox{for all } \Phi\in C^p(A),\,\Psi \in C^{p+1}(A)
.
\eean
One finds
\bean
\delta^T (X^{a_1}
\otimes \ldots \otimes X^{a_ {p+1}})=
\sum_{k=1}^p(-1)^{k+1}X^{a_1}\otimes\ldots\otimes\delta^T (X^{a_k}\otimes X^{a_{k+1}})\otimes \ldots \otimes X^{\alpha_{p+1}}
\eean
where
\bean
\delta^T 
(X^a \otimes  X^b)
=- X^{a+b}+\delta^a_0X^b+X^a \delta^b_0.
\eean 


Let $\delta^{(1)}$ be the restriction of $\delta$ on $C^1(A)$ and 
$$
uX^a=
\left \{
\begin {array}{rcl} 
-\frac 1 {\nu (a)}X^a,&\phantom {h} & a \notin L_0;\\
0,\phantom {X^a}&\phantom {k} &  a \in L_0,\\
\end{array}
\right.
$$
where $$\nu(a)=\mathop{ {\sum}'}_{s=0}^a1=\prod_{i=1}^n(a^i+1)-2.$$
Then $\delta^{(1)+}=u\delta^{(1)T}$ is the generalized inverse of $\delta^{(1)}.$

The space $C^p(A)$ can be decomposed as
\bean
C^p(A)={Ker\, \delta}\oplus {(Ker\, \delta)^\bot},\qquad {Ker\, \delta}={Im\, \delta}\oplus {H^p},  
\eean
where the corresponding orthogonal projectors are given by
\bean
P_{Ker\, \delta}=(I- \delta^+\delta),\quad P_{(Ker\, \delta)^\bot}=\delta^+\delta,\quad P_{Im\, \delta}=\delta\delta^+,\quad 
P_{H^\bullet}=I-\delta^+\delta-\delta\delta^+.  
\eean
The space $H^p=P_{H^\bullet}C^p(A)$ is the Hochschild cohomology group.
From (\ref {tih}) it follows that  $\delta^+$ is nilpotent: \bean (\delta^+)^2= 0.\eean

Let 
\bea \label {lino}
\delta \Phi = \Psi
\eea
be an equation where $\Psi \in C^{p+1}(A)$ is a given cochain and $\Phi \in C^p(A)$ is an unknown cochain. 
Solutions to this equation can be described by a straightforward generalization of the finite-dimensional case
\cite {G}. The space $C^{p+1}(A)$ splits as 
\bean
C^{p+1}(A)={Im\, \delta}\oplus {(Im\, \delta)^\bot},\quad P_{(Im\, \delta)^\bot}=I- \delta\delta^+.  
\eean
Equation (\ref {lino}) has a solution iff  $\Psi\in {Im\, \delta},$ or equivalently,
\bea \label {nesu} P_{Im\, \delta}{\Psi}=\Psi.
\eea                                                    
Then ${\Phi^0} =\delta^+{\Psi}$ is  a specific solution to equation (\ref {lino}) and the general solution is given by \bea \label {geno}
 {\Phi} =\delta^+{\Psi}+ {\Upsilon},
\eea           
where $\Upsilon$ is an arbitrary $p$-cocycle
\bean \label {gen}
 {\Upsilon} =P_{Ker\, \delta} {\Gamma},\qquad \Gamma \in C^p(A).
\eean           

\section {Deformation quantization}

Substituting (\ref {ore}) into (\ref {asse}) one  obtains the system of equations 
\bea \label {sysem}
\delta \Pi^k = \Omega^k,\quad k=1,2,\ldots
\eea
where $\Omega^{k}\in C^3(A)
$ is given by
\bea \label {ors}
\Omega^{k}(f,g,h)=\sum_{m=1}^{k-1}
X^a(X^c f
\Pi^{m}_{c d} X^d g) \Pi^{k-m}_{a b}X^b h-X^a f 
\Pi^{m}_{a b} X^b (X^c g \Pi^{k-m}_{c d}X^d h)
. 
\eea
Since the cochains $\Omega^{k}
$ 
involve only the functions $\Pi^{m}_{a b },m< k,$ and every Poisson manifold has a star product \cite {K}, the solution to system (\ref {sysem}) can be constructed by induction.

If for $\Omega^{k}$ equation (\ref {nesu}) holds: 
\bea \label {nes}  \delta \delta^+{\Omega^k}=\Omega^k,
\eea                                         
then the general solution to 
equation 
(\ref {sysem}) is given by  
\bea \label {gener}
\Pi^k = \delta^+\Omega^k+ (I-\delta^+\delta)\Gamma^k,
\eea
where $\Gamma^k$ is an arbitrary 2-cochain.

The space $C^2(A)$ splits  as 
$$C^2(A)=\tilde C^2(A)\oplus \tilde C^{2 \bot }(A),$$
where
\bea \label {decu} \tilde C^2(A)=\sum_{l\in L}\tilde C^2_l(A),\qquad \tilde C^{2 \bot }(A)=\sum_{l\in L}\tilde C^{2 \bot }_l(A).
\eea
One can show that elements of $\tilde C^{2 \bot }(A)$ are removed from (\ref {ore}) by a similarity transformation and 
the star product can be written as 
\bean \label {or}
f\ast g=fg
+\sum_{k=1}^\infty\sum_{l\in L}t^k
\Pi^k_l
(f,
g)
,
\eean 
where $\Pi^k_l\in \tilde C^2_l(A).$

Let $D$ be the diagonal of $\delta^T\delta.$
The restriction of $\delta^T\delta$  on $\tilde C^2(A)$ is given by
\bean
\delta^T\delta = D - U
\eean
where 
\bean
D(X^a\otimes X^b)=\left(\nu(a)+\nu(b)\right)X^a\otimes X^b, \\ 
U(X^a\otimes X^b)=\mathop{ {\sum}'}_{s=0}^a X^{a-s}\otimes X^{b+s}
+ \mathop{ {\sum}'}_{s=0}^b X^{a+b-s}\otimes X^s
\eean
Elements of the matrix $D$ are positive.All the other elements of $ \delta^T\delta$ are $-1$ or $0$ and the sum of elements of each row equals to $0.$This means that for $\alpha \ne 0$ 
the matrix $\alpha^2 I+\delta^T\delta $
is strictly diagonally dominated.One gets 
\bean
\left(\alpha^2 I+\delta^T \delta \right)^{-1}= (K-U)^{-1}= K^{-1}(I-UK^{-1})^{-1}=
K^{-1}\sum_{m=0}^\infty \left(UK^{-1}\right)^m 
\eean
where $K = \alpha^2 I+D$ is a diagonal matrix. 
Therefore on  $\tilde C^3(A)$ the operator $\delta^+$ is given by an infinite series
\bean
\delta^+= \lim_{\alpha \to 0}
K^{-1}\sum_{m=0}^\infty \left(UK^{-1}\right)^m \delta^T .
\eean

Every equation (\ref {sysem}) splits into a family of independent equations
\bea
\delta \Pi^k_l = \Omega^k_l,\qquad l\in L.
\eea
Let $\tilde C^{2(s)}_l(A)\subset \tilde C^2_l(A)$ and $\tilde C^{2(a)}_l(A)\subset \tilde C^2_l(A)$ be subspaces of symmetric and antisymmetric 2-cochains respectively.Then $\tilde C^2_l(A)=\tilde C^{2(s)}_l(A) \oplus \tilde C^{2(a)}_l(A)$ is an orthogonal decomposition and 
$$\delta  \tilde C^{2(s)}_l(A)\subset \tilde C^{3(a)}_l(A),\qquad \delta  \tilde C^{2(a)}_l(A)\subset \tilde C^{3(s)}_l(A),$$
where $\tilde C^{3(a)}_l(A)\subset \tilde C^3_l(A)$ and $\tilde C^{3(s)}_l(A)\subset \tilde C^3_l(A)$ are flip antisymmetric and flip symmetric 3-cochains respectively. 

For $l=e_i+e_j+e_k,i\ne j,i\ne k,j\ne k,$ define
$$
X^{ij,\,k}=\frac 1 2 (X^{e_i+e_j}\otimes X^{e_k}+ X^{e_k}\otimes X^{ e_i+e_j}),$$
$$
X^{ijk}=\frac 1 2 (X^{e_i}\otimes X^{ e_j}\otimes X^{e_k}- X^{e_k}\otimes X^{ e_j}\otimes X^{e_i}).$$
 The cochains $(X^{ij,\,k},X^{jk,\,i},X^{ki,\,j})$ and  
$(X^{ijk},X^{jki},X^{kij})$ are bases of $\tilde C^{2(s)}_{e_i+e_j+e_k}(A)$ and $\tilde C^{3(a)}_{e_i+e_j+e_k}(A)$ respectively.  From (\ref {one}) and (\ref{ors}) it follows
\bean \label {Om}
\Omega^2_{e_i+e_j+e_k}=\rho_{ijk}+\rho_{jki}+\rho_{kij},\quad
\rho_{ijk}=
\frac 1 {4} \partial^m \omega_{ik}\omega_{mj} X^{ijk},
\eean
where $\partial^m=X^{e_m}.$
One finds 
\bean \label {fin}
\delta^+ X^{ijk}=\frac {1} {3} \left(X^{jk,\,i}-X^{ij,\,k}\right).
\eean
Equation (\ref  {nes}) for $\Omega^2_{e_i+e_j+e_k}$ 
satisfies due to the Jacobi identity 
$$
\partial^m \omega_{ik}\omega_{mj}+\partial^m \omega_{kj}\omega_{mi}+\partial^m \omega_{ji}\omega_{mk}=0.
$$
Using 
(\ref {gener})
one obtains  
\bean \label {assop}
\Pi^2_{e_i+e_j+e_k}(f,g)=\phi_{ijk}(f,g) + \phi_{jki}(f,g)+\phi_{kij}(f,g)+ \mu \delta X^{e_i+e_j+e_k}(f,g),
\eean
where
\bean
\phi_{ijk}(f,g)=\frac 1 {24} (\partial^l \omega_{ij}\omega_{lk}+\partial^l \omega_{ik}\omega_{lj}) \left( \partial^i f \partial^j \partial^k g + \partial^j\partial^k f \partial^i g\right)  
\eean
and $\mu=\mu(x)$ is an arbitrary function.
This expression 
is in agreement with that of refs. \cite {K,PV}.

\bigskip


\end{document}